\documentclass[a4paper,fleqn]{cas-sc}

\usepackage{xcolor}
\usepackage[numbers]{natbib}

\usepackage{pgfplots}
\usepackage{amsmath}
\usepackage{amsfonts}
\usepackage{epstopdf}
\usepackage{lipsum}
\usepackage{fullpage}
\usepackage{tikz}
\usepackage{multirow}
\usetikzlibrary{shapes.geometric}
\usetikzlibrary{arrows.meta,arrows}
\usepackage{algorithm2e}
\usepackage{siunitx}

\newcommand\vecu{\mathbf{u}}
\newcommand\vecS{\mathbf{S}}

\newcommand\VF{\varepsilon_\mathrm{p}}
\newcommand\rbar{\bar{r}}

\usepackage{soul}

\usepackage{hyperref}
\usepackage{cleveref}
\crefformat{section}{\S#2#1#3} 
\crefformat{subsection}{\S#2#1#3}
\crefformat{subsubsection}{\S#2#1#3}

\usepackage{subcaption}

\usepackage{array}

\usepackage[colorinlistoftodos,prependcaption,textsize=small]{todonotes}
\usepackage{xargs}
\newcommandx{\note}[2][1=]{\todo[linecolor=red,backgroundcolor=red!3,bordercolor=red,inline,caption={},#1]{#2}}

\usepackage{lineno}   
\usepackage{amsmath}  
\usepackage{etoolbox} 

\newcommand*\linenomathpatch[1]{%
  \cspreto{#1}{\linenomath}%
  \cspreto{#1*}{\linenomath}%
  \csappto{end#1}{\endlinenomath}%
  \csappto{end#1*}{\endlinenomath}%
}

\linenomathpatch{equation}
\linenomathpatch{gather}
\linenomathpatch{multline}
\linenomathpatch{align}
\linenomathpatch{alignat}
\linenomathpatch{flalign}

\usepackage{adjustbox} 

\graphicspath
{ 
  {actual_figures}
} 

\makeatletter
\def\ps@first{%
   \let\@oddhead\@empty
   \let\@evenhead\@empty
   \def\@oddfoot{}
   \let\@evenfoot\@oddfoot
}

\begin{document}

\title[mode=title]{Drag modelling for flows through assemblies of spherical particles with machine learning: A comparison of approaches}

\author[inst2]{Julia Reuter}
\author[inst1]{Hani Elmestikawy}
\author[inst2,inst3]{Sanaz Mostaghim}
\author[inst1]{Berend van~Wachem\corref{cor1}}[type=editor,
      orcid=0000-0002-5399-40750,
      ]
\cortext[cor1]{Corresponding author: Berend van Wachem, Email: berend.van.wachem@multiflow.org}

\affiliation[inst1]{organization={Chair~of~Mechanical~Process~Engineering},
            addressline={Otto-von-Guericke-Universität},
            city={Magdeburg},
            postcode={39106}, 
            country={Germany}}

\affiliation[inst2]{organization={Chair of Computational Intelligence},
            addressline={Otto-von-Guericke-Universität}, 
            city={Magdeburg},
            postcode={39106}, 
            country={Germany}}

\affiliation[inst3]{organization={Fraunhofer Institute for Transportation and Infrastructure Systems IVI},
            city={Dresden},
            postcode={01069}, 
            country={Germany}}


\begin{abstract}
\noindent Drag forces on particles in random assemblies can be accurately estimated through particle-resolved direct numerical simulations (PR-DNS). Despite its limited applicability to relatively small assemblies, data obtained from PR-DNS has been the driving force for the development of drag closures for much more affordable simulation frameworks, such as Eulerian-Lagrangian point particle methods.  Recently, more effort has been invested in the development of deterministic drag models that account for the effect of the structure of the particle assembly. Current successful deterministic models are mainly black-box neural networks which: 1) Assume pairwise superposition of the neighbours' effect on the drag, and 2) Are trained on PR-DNS data for a wide range of particle concentrations and flow regimes. To alleviate the black-box nature of neural networks, we use genetic programming (GP) to develop interpretable models. In our previous research, this has been proven successful in the Stokes regime. In the current contribution, we extend the application of GP to higher particle Reynolds number regimes. This is done by training a graph neural network (GNN) on the PR-DNS data to learn the pairwise interactions among the particles that constitute the drag variation. The significance of the input features of the GNN is assessed via a feature permutation approach. Then, the estimated pairwise interactions as extracted from the GNN are fed to a GP algorithm, which searches for symbolic expressions that fit the input data. A comparison between the trained GNN model and the resulting symbolic expressions is presented, to assess whether the symbolic expression can capture the underlying patterns learnt by the GNN. The comparison demonstrates the potential of GP in finding relatively simple symbolic models.
At the same time, the accuracy of the symbolic models slightly fall behind the GNN. 
\end{abstract}




\maketitle


\section{Introduction}
\label{sec:intro}
Understanding and simulating gas-solid flow systems relies heavily on the prediction of drag experienced by particles in particle-laden flows~\cite{Balachandar2020a}. Although particle-resolved direct numerical simulations (PR-DNS) can accurately describe particle-fluid interactions and provide high-fidelity flow data, their excessive computational cost limits their use to small-scale systems. However, the data from PR-DNS has driven the development of drag closure models for point particle simulations, which enables more affordable simulations of relatively larger-scale systems.~\par
Recent efforts have focused on deterministic drag models that account for particle arrangement effects, which assume pairwise superposition of neighbour effects~\citep{Akiki2017a}. State-of-the-art deterministic drag models are mostly data-driven and take the form of a black-box neural network providing accurate drag estimation; however, they lack interpretability~\cite{Seyed-Ahmadi2022, Siddani2023, Siddani2024}. Previous work has demonstrated the potential of symbolic regression to develop interpretable drag models in the Stokes Regime. By adopting graph neural networks (GNNs) that apply the pairwise interaction by construction, symbolic regression was able to transfer these pairwise particle interactions into symbolic expressions~\cite{Reuter2023, Elmestikawy2024}. This approach not only provides transparent models, but also helps to understand the underlying physics of particle-laden flows. Building on these findings, the current work seeks to extend this framework to more complex flow regimes, mainly higher Reynolds numbers, where the interplay of inertial effects are more prominent.\par
This paper is organized as follows: Section~\ref{sec:background} provides a background on the particle-laden flow problem at hand, the development of deterministic drag models, and a comprehensive overview of GNN and symbolic regression. Section~\ref{sec:MLframework} details the implemented machine learning framework, including the GNN architecture and the GP algorithm. Section~\ref{sec:results} presents the results of different GNN and GP experiments, comparing the performance of the GNN and the derived symbolic models in various flow regimes. Finally, the conclusions are presented in Section~\ref{sec:conclusion}, with a summary of the key findings.
%
%
\section{Background}
\label{sec:background}

\subsection{Simulation of particle-laden flows}
\label{sec:data_gen}
The PR-DNS data are generated by solving the governing equations of the flow past a periodic arrangement of stationary monodispersed particles, randomly positioned in a cubic domain. The incompressible Newtonian fluid flow is governed by the continuity and the Navier-Stokes equations as follows:
\begin{align}
    \nabla\cdot\vecu &= 0,\\
    \rho \frac{\partial \vecu}{\partial t} + \rho \pmb{\nabla} \cdot (\vecu \otimes \vecu) &= -\nabla p + \mu \nabla^2 \vecu + \vecS_\mathrm{body} + \vecS_\mathrm{IBM},    
\end{align}
where $\rho$ and $\mu$ and the fluid density and the fluid dynamic viscosity, respectively. $\vecu$ is the fluid velocity vector and $p$ represents the pressure. The flow is driven by a body force, $\vecS_\mathrm{body}$, in the direction of the main flow. To enforce the boundary conditions of no-slip and no-penetration at the surfaces of the particles, a momentum source, $\vecS_\mathrm{IBM}$, is introduced. This source term is computed using the hybrid immersed boundary method~\cite{Cheron2023a}.\par
\begin{figure}
    \centering
    \includegraphics[width=0.95\linewidth]{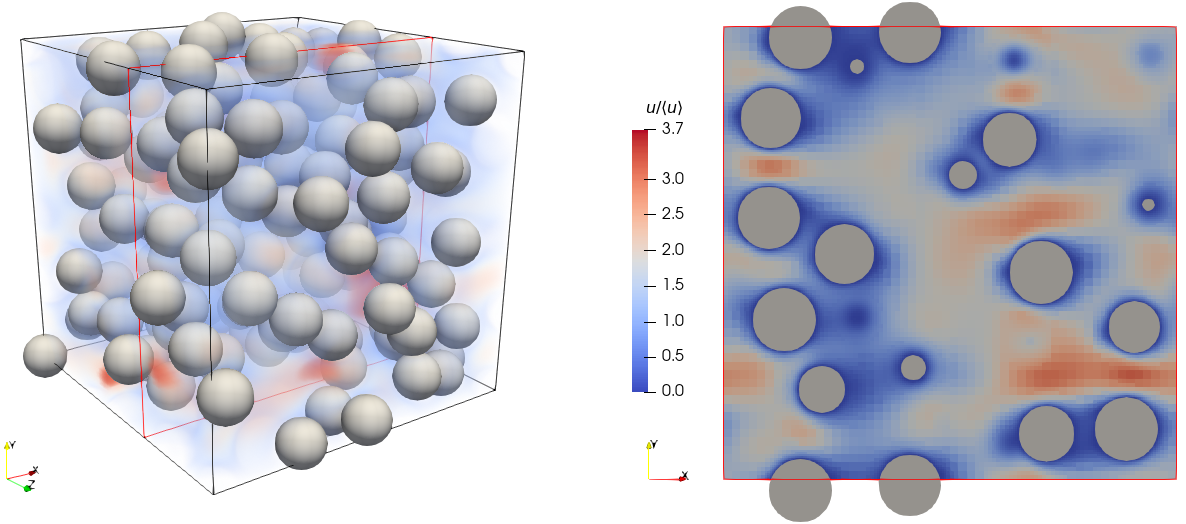}
    \caption{Sample flow field visualizations: 3D (left) and 2D (right), at $\mathrm{Re_p}=5$ and $\varepsilon_\mathrm{p}=0.2$. The flow is driven in the positive $x$ direction.}
    \label{fig:flow_viz}
\end{figure}

The two key global parameters affecting the drag are the particle Reynolds number, $\mathrm{Re_p}$, and the global particle volume fraction, $\varepsilon_\mathrm{p}$. $\mathrm{Re_p}$ is defined based on the fluid properties, particle diameter $\mathrm{d_p}$, and the magnitude of the fluid superficial velocity, $\lVert \langle  \vecu \rangle \rVert$, as follows:
\begin{align}
    \mathrm{Re_p} = \frac{\rho \lVert \langle \vecu \rangle \rVert d_\mathrm{p}}{\mu}.
\end{align}
The governing equations are discretized and solved numerically using a finite volume framework \cite{Denner2020}. The drag force acting on each particle in the arrangement can then be estimated from the discretized solution. Fig.~\ref{fig:flow_viz} shows a sample of the velocity field from one of the considered cases. A dataset is built by varying $\mathrm{Re_p}$, $\varepsilon_\mathrm{p}$, and the random particle arrangement.\par
The considered datasets, originally generated in \cite{vanwachem2024} and are openly available~\cite{vanwachem2023}, cover a range of $\mathrm{Re_p}\in\{0.1, 1, 10, 50, 100, 300\}$ and $\varepsilon_\mathrm{p}  \in \{0.1, 0.2 \dotsc 0.6 \}$. In this work, the datasets are augmented with two additional values of $\mathrm{Re_p}$: $5$ and $200$. Table~\ref{tab:dset_stats} summarizes the mean drag, $\langle F_d \rangle$, and the standard deviation of the drag, $\sigma_d$, present in the datasets across different $(\mathrm{Re_p},\varepsilon_\mathrm{p})$ pairs. The reported mean drag is normalized by the Stokes drag, $F_\mathrm{St} = 3\pi\mu d_\mathrm{p}\lVert\langle \vecu \rangle\rVert$. There exists an extensive body of research that has successfully developed accurate models for the mean drag as a function of $\mathrm{Re_p}$ and $\varepsilon_\mathrm{p}$, e.g., \cite{Richardson1954a,Beetstra2007a,Tenneti2011,Tang2015}. However, these models do not account for the variation of the drag experience by each particle around that mean value~\cite{vanwachem2024}. This variation, as shown in Table~\ref{tab:dset_stats}, can be as high as $\sim 30\%$ of the mean drag force \cite{Akiki2016a}. The primary focus of this work is on the variation of drag forces within the same assembly of particles, i.e., at a given $(\mathrm{Re_p},\varepsilon_\mathrm{p})$ pair, assuming the availability of a perfect mean drag model.
\begin{table}[]
\caption{Statistics of the considered PR-DNS datasets}
\begin{tabular}{c|cc|cc|cc}
\toprule
 & \multicolumn{2}{c|}{$\varepsilon_\mathrm{p}=0.1$} & \multicolumn{2}{c|}{$\varepsilon_\mathrm{p}=0.2$} & \multicolumn{2}{c}{$\varepsilon_\mathrm{p}=0.3$} \\ \midrule
$\mathrm{Re_p}$ & \multicolumn{1}{c|}{$\langle F_d \rangle / F_\mathrm{St}$} & $\sigma_d/\langle F_d \rangle$ & \multicolumn{1}{c|}{$\langle F_d \rangle / F_\mathrm{St}$} & $\sigma_d/\langle F_d \rangle$ & \multicolumn{1}{c|}{$\langle F_d \rangle / F_\mathrm{St}$} & $\sigma_d/\langle F_d \rangle$ \\ \midrule
$0.094$ & \multicolumn{1}{c|}{$2.772$} & $22.78$ & \multicolumn{1}{c|}{$5.668$} & $18.12$ & \multicolumn{1}{c|}{$10.757$} & $20.72$\\
$0.949$ & \multicolumn{1}{c|}{$2.804$} & $23.00$ & \multicolumn{1}{c|}{$5.708$} & $18.25$ & \multicolumn{1}{c|}{$10.828$} & $20.79$\\
$4.863$ & \multicolumn{1}{c|}{$3.152$} & $25.09$ & \multicolumn{1}{c|}{$6.154$} & $19.41$ & \multicolumn{1}{c|}{$11.504$} & $21.46$\\
$9.605$ & \multicolumn{1}{c|}{$3.603$} & $26.42$ & \multicolumn{1}{c|}{$6.809$} & $20.84$ & \multicolumn{1}{c|}{$12.553$} & $22.46$\\
$47.628$ & \multicolumn{1}{c|}{$6.250$} & $26.89$ & \multicolumn{1}{c|}{$11.144$} & $25.06$ & \multicolumn{1}{c|}{$19.860$} & $26.16$\\
$95.304$ & \multicolumn{1}{c|}{$8.831$} & $26.58$ & \multicolumn{1}{c|}{$15.520$} & $26.80$ & \multicolumn{1}{c|}{$27.583$} & $27.95$\\
$188.415$ & \multicolumn{1}{c|}{$13.914$} & $26.70$ & \multicolumn{1}{c|}{$23.773$} & $28.64$ & \multicolumn{1}{c|}{$41.912$} & $29.85$\\
$281.933$ & \multicolumn{1}{c|}{$18.423$} & $26.73$ & \multicolumn{1}{c|}{$31.384$} & $28.70$ & \multicolumn{1}{c|}{$55.436$} & $29.59$\\ \bottomrule
\end{tabular}

\bigskip
\noindent
\begin{tabular}{c|cc|cc|cc}
\toprule
 & \multicolumn{2}{c|}{$\varepsilon_\mathrm{p}=0.4$} & \multicolumn{2}{c|}{$\varepsilon_\mathrm{p}=0.5$} & \multicolumn{2}{c}{$\varepsilon_\mathrm{p}=0.6$} \\ \midrule
$\mathrm{Re_p}$ & \multicolumn{1}{c|}{$\langle F_d \rangle / F_\mathrm{St}$} & $\sigma_d/\langle F_d \rangle$ & \multicolumn{1}{c|}{$\langle F_d \rangle / F_\mathrm{St}$} & $\sigma_d/\langle F_d \rangle$ & \multicolumn{1}{c|}{$\langle F_d \rangle / F_\mathrm{St}$} & $\sigma_d/\langle F_d \rangle$ \\ \midrule
$0.094$ & \multicolumn{1}{c|}{$21.593$} & $18.38$ & \multicolumn{1}{c|}{$42.763$} & $15.71$ & \multicolumn{1}{c|}{$114.182$} & $19.59$\\
$0.949$ & \multicolumn{1}{c|}{$21.744$} & $18.58$ & \multicolumn{1}{c|}{$43.062$} & $15.79$ & \multicolumn{1}{c|}{$114.539$} & $19.60$\\
$4.863$ & \multicolumn{1}{c|}{$22.901$} & $18.69$ & \multicolumn{1}{c|}{$45.361$} & $15.91$ & \multicolumn{1}{c|}{$123.500$} & $19.50$\\
$9.605$ & \multicolumn{1}{c|}{$24.802$} & $19.37$ & \multicolumn{1}{c|}{$48.623$} & $16.25$ & \multicolumn{1}{c|}{$129.842$} & $19.62$\\
$47.628$ & \multicolumn{1}{c|}{$38.310$} & $22.24$ & \multicolumn{1}{c|}{$74.631$} & $18.56$ & \multicolumn{1}{c|}{$211.314$} & $21.16$\\
$95.304$ & \multicolumn{1}{c|}{$53.684$} & $24.95$ & \multicolumn{1}{c|}{$103.123$} & $20.17$ & \multicolumn{1}{c|}{$311.419$} & $21.63$\\
$188.415$ & \multicolumn{1}{c|}{$83.061$} & $27.01$ & \multicolumn{1}{c|}{$164.673$} & $22.14$ & \multicolumn{1}{c|}{$501.833$} & $23.70$\\
$281.933$ & \multicolumn{1}{c|}{$109.347$} & $27.45$ & \multicolumn{1}{c|}{$215.591$} & $22.50$ & \multicolumn{1}{c|}{$672.010$} & $22.45$\\ \bottomrule
\end{tabular}
\label{tab:dset_stats}
\end{table}
\subsection{Physics-informed machine learning}

In the recent literature, an increased interest in the integration of domain-specific knowledge and constraints into machine learning algorithms can be observed. 
Purely data-driven machine learning~(ML) approaches tend to exhibit certain limitations, such as weak generalization on limited or noisy datasets, lack of interpretability, and difficulties with scalability.
Physics-informed machine learning~(PIML) aims at developing models that reflect expected behaviours and adhere to fundamental physical laws~\cite{karniadakis_physics-informed_2021}.
\citet{raissi_physics-informed_2019} introduced physics-informed neural networks~(PINNs), which incorporate physics-based soft constraints on the loss function of the learning algorithm, usually in the form of partial differential equations~(\cite{raissi_hidden_2020}). 
PINNs demonstrated to be effective for a wide range of problems (e.g.,~\cite{cuomo_scientific_2022}), but exhibited ill-conditioned behaviour for increasing problem complexities, as pointed out in~\cite{krishnapriyan_characterizing_2021}.
\citet{watson_machine_2024} provided a comprehensive review of contemporary approaches for incorporating physical principles into neural networks, extending beyond the framework of PINNs.\par
PIML has been applied to the specific problem of predicting the variation around the mean drag in particle-laden flows in the past. One of the initial attempts to model variations in drag using machine learning was the work of \citet{He2019}. In their work, a neural network is trained with the information of all neighbouring particles to predict the force variation. Subsequent work by \citet{Balachandar2020} revealed limitations in this approach, as the neural network suffered from overfitting and required extensive training data to achieve reasonable accuracy. Addressing these challenges, \citet{Seyed-Ahmadi2022} proposed a physics-inspired architecture that replaced the neural network that takes all neighbour information with a repeated neural network designed to process one neighbouring particle at a time, effectively applying pairwise interactions in the network architecture. This approach demonstrated a significant improvement in accuracy, particularly in terms of generalization to unseen data. Building on this progress, \citet{Siddani2023} extended the framework by incorporating trinary and higher-order interactions, achieving a slight improvement in predictive performance.
\subsection{Graph neural networks}

Graph neural networks (GNNs)~\cite{battaglia_interaction_2016,bronstein_geometric_2017} are specifically designed to handle data which is encoded as graph structures.
They exhibit a well motivated inductive bias for problems that can be formulated as interacting entities, as they frequently appear in science and engineering. 
In this context, objects translate to nodes in a graph, and their pairwise interactions are represented as edges between these nodes. 
\citet{battaglia_relational_2018} provide a generalized GNN framework for the applicability to a wide range of problem domains. 
Their proposed generalized GNN block is based on directed multi-graphs with attributes at different levels.
The definition of a graph is a 3-tuple ${G = (\mathbf{u}, V, E)}$, which we further illustrate using the sun and planets of our solar system as an example.
In this system, each mass body is modelled as a node of the graph, and their mutual gravitational interactions as edges between them.
This example, studied in more detail in~\cite{lemos_rediscovering_2023}, shares similarities with our problem, but differs in terms of scale (planets vs. particles):
\begin{itemize}
    \item $\mathbf{u}$ represents global or universal properties. 
    A universal property in the solar system could be the gravitational constant or the overall gravitational field that influences all celestial bodies.
    \item $V = \{\mathbf{v}_i\}_{i=1:N^v}$ is the set of nodes, which represent celestial bodies such as the Sun and planets.  
    The total number of nodes in the graph is \(N^v\). 
    $\mathbf{v}_i$ are the node features, which describe properties such as mass, position, or velocity of each celestial body in the system.
    \item $E = \left\{ \left(\mathbf{e}_k, r_k, s_k\right) \right\}_{k=1:N^e}$ is the set of edges, describing relations between bodies. 
    $N^e$ represents the number of edges in the graph. 
    $\mathbf{e}_k$ contains the edge features, such as the gravitational force acting between two bodies.
    An edge connects the receiver node with the index $r_k$ to the sender node $s_k$, representing the direction of gravitational influence.
\end{itemize}

Using the terminology introduced in~\cite{battaglia_relational_2018}, a few computations are required to predict a target variable at the node level, considering influences from the connected nodes.
These include two network blocks, namely an edge model $\phi^e$ and a node model $\phi^v$, together with an aggregation function of messages on the node level, denoted by $\rho^{e \rightarrow v}$.
\begin{align}
    \mathbf{e}'_{k} &= \phi^e(\mathbf{e}_{k},\mathbf{v}_{r_k}, \mathbf{v}_{s_k}, \mathbf{u}) &\text{\textit{message computation }} \label{eq:msg_comput}\\ 
    \bar{\mathbf{e}}'_i &= \rho^{e \rightarrow v}(\{ \mathbf{e}'_{k}, r_k, s_k\}_{r_k=i,k=1:N^e} ) &\text{\textit{message aggregation }} \label{eq:msg_agg}\\
    \mathbf{v}_i' &= \phi^v(\bar{\mathbf{e}}'_i,\mathbf{v}_i,\mathbf{u}) &\text{\textit{node update }} \label{eq:node_update}
\end{align}

The node and edge models are comparably small ANN blocks, which are shared among all nodes and edges.
In other words, all messages between two nodes are computed by the same edge model, and all node updates are computed by the same node network.
In Eq.~\ref{eq:msg_comput}, a message is computed by the edge model, using the edge features, the features of the sending and receiving nodes as well as the universal property. 
This can be understood as the approximation of pairwise interactions between entities.
All incoming messages of a node are aggregated in Eq.~\ref{eq:msg_agg}, requiring an aggregation function invariant to permutations of the input, such as elementwise summation or mean. 
The node update is learned by the node model, using the aggregated messages, the current node state and the universal property as input features. 

Different applications of GNNs to systems concerning fluid mechanics have been presented in the recent literature. 
~\citet{sanchez-gonzalez_learning_2020} have introduced an encoder-decoder architecture based on graphs to model physical processes, including water dynamics, sand friction, and the behaviour of viscous material. 
Additionally, \citet{li_accelerating_2021} has developed two variations of GNNs, one emphasizing edges and the other focusing on nodes, to enhance the efficiency of Lagrangian fluid simulations.
Network-based ML methods, however, typically lack interpretability, which is a significant drawback for model inference tasks, where understanding the underlying principles is as critical as achieving accurate predictions.

To overcome this issue, symbolic regression~(SR) can be used to develop interpretable equations. 
Although the learning capability of SR is generally considered lower than that of network-based models, this method has the inherent advantage of generating mathematical expressions, thus improving the model explainability.
In addition, a well-performing symbolic model can, in certain cases, generalize better and mitigate the extrapolation limitations often observed in ANNs.
\citet{cranmer_discovering_2020} proposed GNNs as an inductive bias for SR, aiming to overcome issues related to high-dimensional problems that can be modelled as interacting particle systems.
In the first step, they approximated the pairwise interactions between particles using GNNs.
Since the edge and node models are shared for all particles and interactions between them, this separable internal structure breaks originally high-dimensional problems into smaller subproblems, which are tractable for SR algorithms.
In the next step, the edge and node models were approximated by an SR algorithm, operating on a reduced amount of input features compared to the original problem. 
This method was applied successfully to rediscover the gravity equation for international trade in~\cite{verstyukMachineLearningGravity2022a}, the collective behaviour of swarms~\cite{powers_extracting_2022}, as well as the orbital mechanics of the solar system according to Newton's gravitation law~\cite{lemos_rediscovering_2023}.
We demonstrated the applicability of this two-step ML pipeline to learn symbolic expressions for the variation from the mean drag on particles in particle-laden flows in two previous studies~\cite{Elmestikawy2024,Reuter2023}. 
While earlier results were presented for the Stokes regime, in the present paper, we extend these findings to a broader range of $\mathrm{Re_p}$ and $\varepsilon_\mathrm{p}$.
%
%
\subsection{Symbolic regression}

\begin{figure}
    \centering
    \begin{tikzpicture}[
  edge from parent/.style={draw,thick,-{Latex[length=3mm]},blue!60},
  every node/.style={draw,thick,rounded corners,rectangle,inner sep=3pt,fill=blue!15,draw=blue,},
  level 1/.style={sibling distance=25mm},
  level 2/.style={sibling distance=15mm},
  level distance=1.1cm
  ]
\node { $\; \cdot \;$}
  child { 
    node { $+$ }
      child { node { $a$ } }
      child { node { $3.8$ } }
  }
  child { 
    node { $-$ }
      child { node { $\cos$ }
        child { node { $z$ } }
      }
      child { node { $\; \cdot \;$ }
        child { node { $1.4$ } }
        child { node { $b$ } }
      }
  };

\node [right=0.3cm of current bounding box.east, align=center] {
  Symbolic Expression: \\
  $(a + 3.8) \cdot \cos(z) - (1.4 \cdot b)$
};

\end{tikzpicture}
    \caption{Representation of a symbolic expression as a syntax tree.}
    \label{fig:gptree}
\end{figure}
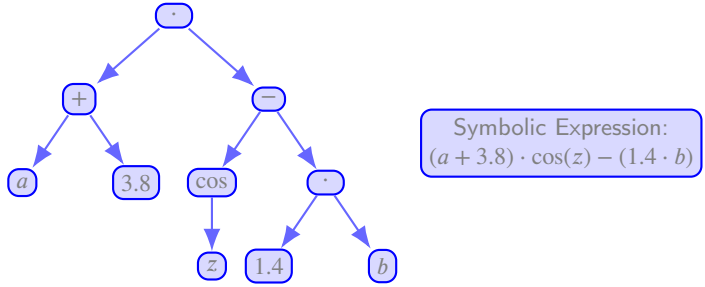

Symbolic regression learns symbolic expressions from data and, other than classical regression methods, does not assume a specific functional form a priori. 
The identification of a fitting model structure is part of the learning task, next to learning the coefficient values and selecting appropriate functions and features from the provided sets.
Various techniques have been presented in the past to approach SR problems. 
A popular method for physics and fluids is sparse identification of nonlinear dynamical systems~(SINDy), which exploits sparse regression to describe underlying principles of nonlinear dynamical systems~\cite{brunton_discovering_2016,de_silva_discovery_2020}. 
AI Feynman~\cite{udrescu_ai_2020,udrescuAIFeynmanParetooptimal2020} aims at identifying functions for physical systems, which often share characteristics such as symmetries and separability, and require the adherence of physical units.
Recent developments in the area of SR are based on artificial neural networks (ANNs) to learn a symbolic model.
Deep symbolic regression (DSR)~\cite{petersen_deep_2021} formulates SR as a reinforcement learning problem with an underlying recurrent neural network to generate equations as a sequence of tokens. 
\citet{tenachi_deep_2023} introduced a DSR approach for physics, incorporating unit constraints to guide the learning process.
Dimensional analysis was applied as in situ constraints, which updated the probability distributions for the token selection to ensure unit-consistent operations and features.
The area of ANN-based SR is currently emerging rapidly, with many new approaches frequently presented that are of relevance for science and engineering problems.
This encompasses various approaches, such as the dynamic adjustment of the network architecture~\cite{li_metasymnet_2023}, the learning of ordinary~\cite{dascoli_odeformer_2023} and partial differential equations~\cite{brunton_machine_2023}, transformer-based architectures~\cite{kamienny_end--end_2023} as well as deep generative SR~\cite{holt_deep_2023}.

Some physics-inspired SR methods are constrained to particular system types or predefined equation structures, which can hinder their capacity to discover new or unexpected physical laws and limit their flexibility when applied to complex physical systems.
Genetic programming (GP) is a long-standing method in the literature and can be considered a flexible approach for SR.
It allows for extensive customization, enabling the algorithm to evolve a wide variety of equation forms and to adapt to different problem domains.
Therefore, GP is the preferred method for SR in the approach presented in this paper. In the following, we will briefly introduce the fundamentals of GP and then focus on recent developments in this area, that are relevant for our approach. 

GP is a method from the family of evolutionary algorithms, which operates on expressions that are represented as syntax trees (see Fig.~\ref{fig:gptree}).
GP formally requires a set of terminals, denoted by $\mathcal{T}$, which are features or constants located at the leaf nodes of the tree, and a set of functions, denoted by $\mathcal{F}$, which are applied to these terminals.
Being a population-based algorithm, GP evolves multiple expression simultaneously, which are combined and refined throughout evolution to converge to expressions that can accurately predict the target feature.
A basic GP algorithm with an additional archive of non-dominated solutions according to Pareto dominance is outlined in Algorithm~\ref{algo:canonicalGP}.


\SetStartEndCondition{ }{}{}\SetKwProg{Fn}{function}{}{end}
\SetKwFunction{choice}{choice}
\SetKwComment{Comment}{/* }{ */}
\SetKwInOut{Input}{Input}
\SetKwInOut{Output}{Output}
\SetStartEndCondition{ }{}{}
\SetKwProg{Fn}{function}{}{end}
\SetKwProg{Fn}{function}{\string:}{}
\RestyleAlgo{ruled}
\DontPrintSemicolon
\newcommand\notterm{\textnormal{\textbf{not} termination criterion$\;$}}
\newcommand\iinlambda{$i = 1\dots \lambda \;$}
\newcommand\nondominds{\textnormal{non-dominated individuals from $P(t) \;$}}
\begin{algorithm}[t]
\DontPrintSemicolon

\Input{Training data $X$, population size $\mu$, number of children $\lambda$, crossover probability $p_c$, mutation probability $p_m$, maximum generations $G$}
\Output{Solution population $P$}
  $A \gets \emptyset$  \tcp*{empty Pareto-dominance based archive}
  $P \gets $ $\textnormal{initializePopulation}(\mu)$ \tcp*{create random initial population}
  
  $\textnormal{evaluate}(P,\, X)$ \tcp*{compute objective values}
  
  \For{$G$\textnormal{ generations\,}}{
    $P_{\mathrm{offspring}} \gets \emptyset$ \tcp*{create empty child population}
    \For{\iinlambda}{
        $genOp \gets \textnormal{selectGenOp}(p_c, \, p_m)$ \tcp*{select genetic operator}
        $p \gets \textnormal{select}(P)$ \tcp*{select parent(s)}
        $o \gets \textnormal{reproduce}(p,genOp)$ \tcp*{create offspring}
        $P_{\mathrm{offspring}} \gets P_{\mathrm{offspring}} \cup o$ \tcp*{add to offspring population}
        }
    $\textnormal{evaluate}(P_{\mathrm{offspring}}, \, X)$ \tcp*{evaluate offspring population}
    $P \gets \textnormal{updatePopulation}(P \cup P_{\mathrm{offspring}})$  \tcp*{update population}
    $A \gets \textnormal{updateArchive}(A \cup P_{\mathrm{offspring}})$  \tcp*{update non-dominated archive}

  }
\Return $A$ \tcp*{return Hall of Fame}
\caption{Genetic Programming Algorithm}
\label{algo:canonicalGP}
\end{algorithm}

\begin{figure}[t]
    \centering
        \begin{subfigure}[b]{0.9\textwidth}
    \centering
    \resizebox{11cm}{!}{
\begin{tikzpicture}[
    edge from parent/.style={draw,thick,-{Latex[length=3mm]},blue!60},
    every node/.style={draw,thick,rounded corners,rectangle,inner sep=3pt,fill=blue!15,draw=blue,},
    level 1/.style={sibling distance=25mm},
    level 2/.style={sibling distance=15mm},
    level distance=1.1cm
    ]
    \node (p1root) {$\;\cdot\;$}
      child {node (p1a) {$-$}
        child {node (p1a1) {$a$}}
        child {node (p1a2) {$2.0$}}
      }
      child {node (p1b) {$\cos$}
        child {node (p1b1) {$z$}}
      };
    \node[above=of p1root, node distance=0.2cm] {Parent 1};
    \draw[red,thick] (p1root) edge[-{Latex[length=3mm]},red] (p1b);
    \node[right=3.5cm of p1root] (p2root) {$\cos$}
    child {node (p2) {$\; \cdot \;$} 
          child {node (p2a) {$5.0$}}
          child {node (p2b) {$w$}}
    };
    \node[above=of p2root, node distance=0.2cm] {Parent 2};
    \draw[red,thick] (p2root) edge[-{Latex[length=3mm]},red] (p2);   
    \node[right=4.2cm of p2root] (o2root) {$\;\cdot\;$}
      child {node (o2a) {$-$}
        child {node (o2a1) {$a$}}
        child {node (o2a2) {$2.0$}}
      }
      child {node (p2) {$\;\cdot\;$} 
          child {node (p2a) {$5.0$}}
          child {node (p2b) {$w$}}
        };
    \node[above=of o2root, node distance=0.2cm] {Offspring};
\end{tikzpicture}
}
\caption{One-point Crossover}
\label{fig:crossover}
\vspace*{5mm}
\end{subfigure}

\begin{subfigure}[b]{0.9\textwidth}
   \centering 
\resizebox{8cm}{!}{
\begin{tikzpicture}[
    edge from parent/.style={draw,thick,-{Latex[length=3mm]},blue!60},
    every node/.style={draw,thick,rounded corners,rectangle,inner sep=3pt,fill=blue!15,draw=blue,},
    level 1/.style={sibling distance=25mm},
    level 2/.style={sibling distance=15mm},
    level distance=1.1cm
    ]
    \node (p1root) {$\;\cdot\;$}
      child {node (p1a) {$-$}
        child {node (p1a1) {$a$}}
        child {node (p1a2) {$2.0$}}
      }
      child {node[draw,thick,rounded corners,rectangle,inner sep=3pt,fill=red!25] (p1b) {$\cos$}
        child {node (p1b1) {$z$}}
      };
    \node[above=of p1root, node distance=0.2cm] {Parent};
    \node[right=5cm of p1root] (o2root) {$\;\cdot\;$}
      child {node (o2a) {$-$}
        child {node (o2a1) {$a$}}
        child {node (o2a2) {$2.0$}}
      }
      child {node (o2b) {$\exp$}
        child {node (o2b2) {$z$}}};
    \node[above=of o2root, node distance=0.2cm] {Offspring};
\end{tikzpicture}
}
\caption{Point Mutation}
\label{fig:mutation}
\end{subfigure}
    \caption{Examples of crossover and mutation operation, where crossover points are marked as red arrows and mutation points as red nodes.}
    \label{fig:reproduction}
\end{figure}
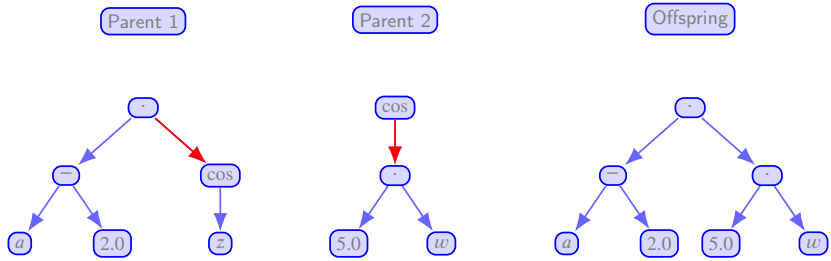
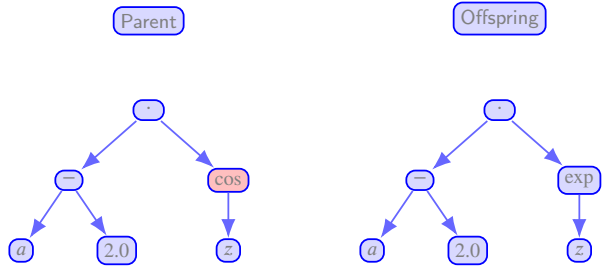

First, an initially random population of expressions is generated using $\mathcal{T}$ and $\mathcal{F}$.
These are then evaluated with respect to the objective functions. 
Afterward, $\lambda$ children are created and added to the offspring population.
In the reproduction step, the evolutionary operators, namely crossover and mutation, are employed. 
These operate on the tree structures of the syntax tree, where crossover exchanges subtrees of two parents to create offspring, and mutation inserts small modifications into the tree. 
Examples for crossover and mutation are given in Fig.~\ref{fig:reproduction}.
The offspring population is evaluated, before the parent population gets updated with the newly generated offspring.
To this end, we employ a $\mu + \lambda$ reproduction scheme, where both, the parent population and all generated offsprings, are considered in the update.
This process is repeated and expressions are refined iteratively until a stopping criterion, typically a maximum number of generations, is reached. 
Next to an error measure, the complexity of the equation is typically minimized to avoid the uncontrolled growth of equations that only have small improvements in terms of accuracy. 
This turns the problem into a multi-objective one, which affects the selection and update-steps of the algorithm.
To this end, the multi-objective NSGA-II algorithm~\cite{deb_fast_2002} is frequently employed, which optimizes multiple objectives simultaneously. 
Rather than a single optimal solution, a Pareto-optimal front of solutions, each satisfying the objectives to different degrees, is returned by the algorithm.
At least one of the two steps, selection of parents and updating of the population, has to be fitness-based to ensure convergence to better solutions over time. 
Fitness-based in this context means that expressions with better objective values are more likely to get selected as parents, or to survive to the next generation. 
Finally, the algorithm returns an archive of non-dominated solutions, from which a decision maker will select one expression with a good trade-off between the objectives.
This archive of the best solutions is typically referred to as the Hall of Fame (HoF).

State-of-the-art  GP algorithms are frequently enhanced by components from the ML area to improve the computational efficiency and accuracy of the developed expressions. 
One important component concerns the fitting of new constants in an expression.
This is typically implemented by an additional constant placeholder in the feature set. 
Before the evaluation of an equation, these placeholder constants are fitted using a multiple linear regression algorithm on top of GP.
For example, the \texttt{PySR}~\cite{cranmer_interpretable_2023} framework employs the Broyden–Fletcher–Goldfarb–Shanno (BFGS) algorithm, and \texttt{TiSR}~\cite{martinek_introducing_2023} and \texttt{Operon}~\cite{burlacu_operon_2020} implement the Levenberg-Marquardt algorithm to fit new constants.

Similar to the area of PIML, the integration of domain-specific knowledge into GP algorithms has become increasingly important in recent research.
By incorporating prior knowledge, expressions can be constrained to follow expected behaviours or patterns, while reducing the search space of possible equations and improving accuracy and interpretability. 
Moreover, this approach can help overcome the limitations of purely data-driven methods, such as lack of generalization and physical consistency.
We will present a few notable publications on this topic subsequently, some of which will also be employed in our proposed approach. 

\label{sec:gp}
One line of research concerns the inclusion of shape constraints in GP algorithms. 
Shape constraints can be useful when prior knowledge about the expected function behaviour is available, such as monotonicity, symmetry or convexity, but the data is not sufficient to fully reflect this behaviour~\cite{kronberger_shape-constrained_2022}.
Analysing the function behaviour using optimistic~\cite{bladek_solving_2019} and pessimistic approaches~\cite{haider_shape-constrained_2022,haider_comparing_2022} has been studied in the literature. 
Moreover, the application of appropriate constraint handling methods was investigated, such as minimizing the constraint violations as an additional objective using multi-objective optimization~\cite{bladek_solving_2019,haider_shape-constrained_2022,moreno-diaz_shape-constrained_2022}, or counter-example driven SR~\cite{bladek_solving_2019,kubalik_symbolic_2020}.

Another important method to include prior knowledge is unit-aware GP, which aims at developing expressions that are conformal with physical units. 
This is especially relevant for problems in science and engineering, where measurements and their associated units are frequently available.
Equations that, for example, add two quantities in meters and seconds are typically not plausible for such applications. 
Unit-aware GP propagates the units of the features through the syntax tree and detects unit violations.
Again, constraint handling methods are required to address potential unit violations. 
These include adding a high penalty value to the primary objective for unit-violating equations~\cite{bandaru_dimensionally-aware_2013,mei_constrained_2017}, multi-objective optimization~\cite{li_dimensionally_2020,zille_assessment_2021}, as well as a repair mechanism to remove unit violations in expressions~\cite{keijzer_dimensionally_1999}.
Another challenge is related to the dimensional analysis of equations that contain new constants, which typically have unknown units.
However, all features and constants require units to be considered in the dimensional analysis and propagated through the tree.
Our recently proposed unit-aware GP approach helps to overcome this issue~\cite{reuter_unit-aware_2024}. 
It treats the units of newly learned constants as jokers in the dimensional analysis and propagates them further through the tree.

Other methods to introduce knowledge into GP algorithms include prior assumptions about the functional structure, such as fixing parts of the function to a known sub-expression~\cite{asadzadeh_symbolic_2021,schwab_improving_2012}, or seeding the initial equations with expected building blocks~\cite{schmidt_incorporating_2009}. 
The pre-computation of expected combinations of functions and features and their inclusion in the feature set improved the accuracy of the final equations in ~\cite{versino_data_2017,zille_assessment_2021}.

For the specific problem of particle-laden flows, our proposed approach builds upon our previous research on the development of symbolic models for particles in Stokes flow using GP.
These include the flow past one~\cite{zille_assessment_2021} and two~\cite{reuter_towards_2022} spherical particles, as well as using GNNs as inductive bias for GP for larger particle arrangements in infinite~\cite{Reuter2023} and periodic~\cite{Elmestikawy2024} flow domains.
The goal in this paper is to assess the performance of the proposed framework on a broader range of Reynolds numbers and volume fractions. 

\section{Machine learning framework}
\label{sec:MLframework}

\subsection{Overall approach}
With the two-step ML pipeline introduced subsequently, we aim to predict the variation from the mean drag force experienced by a particle. 
The mean drag is denoted by $\langle F_x \rangle$, and $\Delta F_{x,i}$ denotes the variation from the mean drag.
Typically, the deviation from the mean drag on a particle is formulated as a function of the positions of a predefined number of neighbours, denoted by $N_n$.

\begin{equation}
    \frac{\Delta F_{x,i}}{\langle F_x \rangle} = f_x(\VF, \mathrm{Re_p}, \rbar_1,\theta_1,\varphi_1, \rbar_2,\theta_2,\varphi_2, \dots, \rbar_{N_n},\theta_{N_n},\varphi_{N_n}).
\end{equation}

Depending on the number of neighbours, typically 20 or more, this problem formulation exhibits high-dimensional properties, which are often difficult to handle for data-driven ML methods.
An initial approach using a fully connected neural network considering all features at the same time showed poor generalization capabilities~\cite{Balachandar2020}, which can be attributed to the problem formulation.
To improve the model generalization, we assume pairwise interactions between particles, which reduces the number of features for each pairwise interaction~\cite{Seyed-Ahmadi2020}. 
This moreover motivates the use of GNNs, as they are designed to model pairwise interactions between particles while preserving the relationships between them.  
Consequently, our target model aggregates the pairwise interactions, where each individual pairwise interaction is approximated by a function $\Tilde{f}_x$.

\begin{equation}
    \frac{\Delta F_{x,i}}{\langle F_x \rangle} = \sum_{j=1}^{N_\mathrm{n}} \Tilde{f}_x(\VF, \mathrm{Re_p}, \rbar_j,\theta_j,\varphi_j).
\end{equation}

A pairwise interaction is thus a function of the volume fraction, the Reynolds number, and the relative distance of the neighbour particle from the particle of interest in spherical coordinates.
The particle of interest is always located at the origin of the coordinate system.

\begin{figure*}[t]
    \centering
    \includegraphics[width=\linewidth]{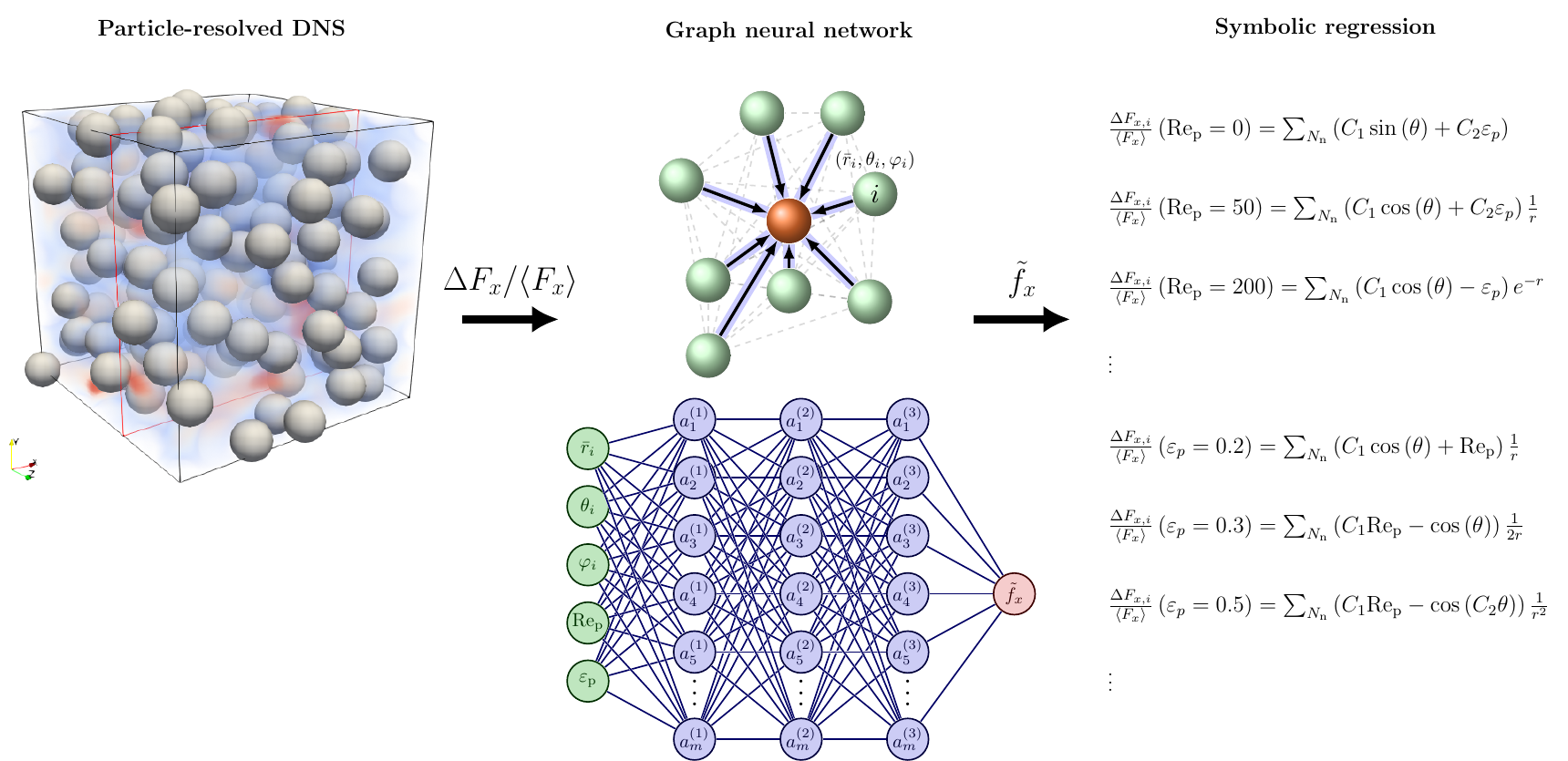}
    \caption{Symbolic models are generated from simulation data using a GNN as a surrogate model. Physical particles in the simulation translate to nodes in the GNN.}
    \label{fig:flowchart}
\end{figure*}

\begin{table}[t]
    \caption{Exemplary intermediate data products as extracted from the GNN. Each row corresponds to one pairwise interaction between a neighbor and the particle of interest. These data serve as training data for the subsequent GP algorithm, with the target variable $\Tilde{f}_x$. }
    \centering
    \begin{tabular}{c|c|c|c|c|c}
    \toprule
         $\rbar_j$ & $\theta_j$ & $\varphi_j$ & $\varepsilon_p$ & $\mathrm{Re_p}$ & $\Tilde{f}_x$\\
         \midrule
         $1.051813$ & $1.998868$ & $0.660557$ & $0.4$ & $188.415$  & $0.033569$ \\
         $1.302048$ & $2.127855$ & $-1.610042$ & $0.4$ & $281.933$  & $0.004412$\\
         $1.857303$ & $0.652636$ & $2.494535$ & $0.4$ & $47.628$ & $-0.011249$\\
         $1.503460$ & $2.399730$ & $2.959001$ & $0.3$ & $4.863$ &  $-0.053079$ \\
         $2.092653$ & $0.131259$ & $3.052247$ & $0.3$ & $281.933$  &  $-0.090326$ \\
         $1.100217$ & $2.058815$ & $2.576300$ & $0.5$ & $95.304$ &   $0.016363$ \\
         $1.975268$ & $0.748523$ & $-2.570214$ & $0.2$ & $188.415$   & $-0.015988$ \\
         $3.202558$ & $2.437589$ & $-1.476869$ & $0.1$ & $0.094$  & $-0.006449$ \\
         $1.005887$ & $1.695564$ & $0.834028$ & $0.4$ & $95.304$  & $0.186315$ \\
         $\cdots$ & $\cdots$ & $\cdots$ & $\cdots$ & $\cdots$ & $\cdots$ \\
\bottomrule

    \end{tabular}
    \label{tab:intermediate_data}
\end{table}

The overall learning pipeline is displayed in Fig.~\ref{fig:flowchart}. 
The GNN is fed with data from the particle-resolved simulations and learns the underlying pairwise interactions between the particle of interest and its neighbours. 
The edge model $\Phi^e$ is shared to this end, so that each pairwise interaction is approximated by the same model. 
The pairwise interactions are aggregated internally to predict the target feature $\frac{\Delta F_{x,i}}{\langle F_x \rangle}$.
In the previously introduced structure of the GNN, these operations correspond to Eqs.~\ref{eq:msg_comput} and~\ref{eq:msg_agg}.
Eq.~\ref{eq:node_update} is not considered in our approach, as our experiments from previous studies have indicated that the accuracy does not improve significantly by an additional node update~\cite{Reuter2023,Elmestikawy2024}.
We employ a single GNN to learn the underlying interactions for all $\mathrm{Re_p}$ and $\VF$.
In this way, we aim to ensure that both the GNN and symbolic models maintain a consistent pattern across the entire range of input data, which is important for their generalizability and reliability.
Once the training phase of the GNN is finished, a dataset with intermediate data products is generated, which includes the input and output features of the edge model $\Phi^e$. 
Examples for these intermediate data products are shown in Tab.~\ref{tab:intermediate_data}.
A GP algorithm uses this data to replace the edge model with symbolic expressions.
Since the training data contains one pairwise interaction per training sample, the number of input features decreases compared to the problem formulated without pairwise interactions. 
While the GNN employed samples from all available $\mathrm{Re_p}$ and $\VF$, we know from previous studies that the GP algorithm has difficulties identifying a universal equation that covers the entire range of data.
Therefore, we filter the dataset and develop separate equations, once for each $\mathrm{Re_p}$ with $\VF$ as a variable, and once for each $\VF$ with $\mathrm{Re_p}$ as a variable.
This reduces the space of searched equations tremendously, and simplifies the learning for GP. 
An important prerequisite to this end is that the GNN has to achieve a certain accuracy; otherwise, errors in the first step will be further propagated to the resulting symbolic expressions. 

The GNN module of our approach generally has a high similarity with the physics-inspired neural network by \citet{Seyed-Ahmadi2020}, which approximates pairwise interactions with small network blocks. 
However, the use of GNNs embedded in the \texttt{PyTorch geometric} library as in our approach allows for easier customization and extension in the future, such as trinary interactions between particles or alternative neighbourhood structures. 

\subsection{Algorithm settings}
\label{sec:algo_settings}
In the following, we will describe the settings of the previously introduced ML pipeline as they are employed in our experiments.
The GNN is trained using three fully connected layers, with rectified linear unit (ReLU) activations in the first two layers and a final hyperbolic tangent activation. 
Using ReLU in the first layers helps to avoid the problem of vanishing gradients, which is a known drawback of the hyperbolic tangent activation.
ReLU moreover allows for sparse activations, since $\mathrm{ReLU}(x) = \max\{0,x\}$, which can be useful for covering diverse datasets like ours, containing various $\mathrm{Re_p}$ and $\VF$.
The final hyperbolic tangent activation constrains the output range, which is meaningful for our problem, as forces are naturally bounded.
In our previous studies~\cite{Reuter2023,Elmestikawy2024}, we trained separate GNNs for each $\VF$ in Stokes flow at $\mathrm{Re_p}=0$ and employed two hidden layers to this end.
Here, we introduced an additional layer to better accommodate the increased number of parameters in the data. 
Generally, the network size must be sufficiently large to capture the underlying interactions between particles, while remaining compact enough to ensure good generalization.
We selected 30 neurons per hidden layer, which we also employed in previous studies~\cite{Reuter2023,Elmestikawy2024} and has shown to be a good compromise between model expressiveness and generalization. 
A dropout probability of 0.05 is applied after the second ReLU layer to prevent overfitting while maintaining stable training dynamics.
Training is conducted using the Adam optimizer with an initial learning rate of 0.002 for 10,000 epochs, with a batch size of 64. 
Each epoch corresponds to a complete pass through the training dataset.
The loss function is the mean-squared error (MSE), which is a useful loss function for regression problems. 
The training features include the relative locations of the neighbouring particles in spherical coordinates, $\rbar_j$, $\theta_j$, and $\varphi_j$, as well as $\mathrm{Re_p}$ and $\varepsilon_\mathrm{p}$.
Our implementation is based on \texttt{PyTorch}~\cite{Paszke2019} for general neural network operations, and \texttt{PyTorch geometric}~\cite{fey_fast_2019} for tasks specific for GNNs. 
Tab.~\ref{tab:gnn_parameters} summarizes these experiment settings. 

\begin{table}[t]
    \centering
    \caption{Parameter settings used to train the GNN}
    \begin{tabular}{l|l}
    \toprule
    GNN parameter & Settings \\
    \midrule
    $\Phi^{e}$ & 3 fully connected layers \\
    Nonlinearity & ReLU — ReLU — hyperbolic tangent \\
    Dropout probability & 0.05, applied after the second ReLU layer\\
    Number of neurons & 30 per hidden layer \\
    Initial learning rate & 0.002 \\
    Training features & $\rbar_j$, $\theta_j$, $\varphi_j$, $\mathrm{Re_p}$, $\varepsilon_\mathrm{p}$ \\
    Target feature & $\frac{\Delta F_{x,i}}{\langle F_x \rangle}$ \\
    Optimizer & Adam \\
    Training duration & 10,000 epochs \\
    Batch size & 64 \\
    Loss function & MSE \\
    Implementation & \texttt{PyTorch}~\cite{Paszke2019}, \texttt{PyTorch geometric}~\cite{fey_fast_2019}\\
    \bottomrule
    \end{tabular}
    \label{tab:gnn_parameters}
\end{table}

The GP algorithm is applied once the training of the GNN is finished, and intermediate data products have been generated. 
We employ a population of 500 expressions, from which 500 children are created in each generation using genetic operators like crossover and mutation. 
From this combined pool of 1000 expressions, the best 500 according to the NSGA-II algorithm survive to the next generation, where this process is repeated for 200 generations. 
We set an upper limit of 30 for the complexity of an equation to avoid bloat and reduce the search space to a reasonable size. 
The optimization objectives applied to the population at the selection step include the MSE as the primary objective, supported by three additional objectives:
First, the Spearman correlation, as previous studies indicated that it can help to keep solutions with high correlation but also high error in the population, and allow for further refinement in the following generations.
Second, a dimension penalty is implemented, which aims to minimize operations that violate physical units. 
We consider the units as displayed in Tab.~\ref{tab:gp_parameters}, where $[-]$ represents dimensionless quantities and $[?]$ is a joker unit as introduced by tuneable constants that come without an assigned unit. 
We want to mention that we assign a unit of $[\unit{\metre}]$ to the distance from the centre particle, $\rbar_j$, even though this feature is provided in a normalized and dimensionless format in the training features.
We do this to reflect the semantics of the feature as a distance in the algorithm and avoid operations such as $\sin(\rbar_j)$.
Such operations would be allowed if $\rbar$ was a dimensionless unit; however, they lack physical meaning from a fluid mechanics perspective. 
More details on our multi-objective unit-aware GP approach, which includes the mentioned joker units in the dimensional analysis, can be found in our previous work~\cite{reuter_unit-aware_2024}.
Lastly, we aim to minimize complexity since simpler solutions typically tend to generalize better. 
This objective enables the simultaneous optimization of MSE across all complexity levels during evolution. 
The Spearman correlation is not included in the HoF, since it is designed to store only those solutions that already perform well in terms of MSE and do not require further refinement. 

Since we develop equations either for each $\VF$ with $\mathrm{Re_p}$ as a variable, or for each $\mathrm{Re_p}$ with $\VF$ as a variable, only one of the two are included in the feature set.
The used set of functions comprises basic arithmetic operations, as well as trigonometric, logarithmic and exponential functions. 
Moreover, a power operator, denoted by $\circ^{\circ}$ is employed, which takes two input arguments, the basis and the exponent, where the latter can include a term consisting of multiple operations. 

Custom complexity values are employed to reflect the complexity of operations in the complexity of the expression. 
While features are assigned a complexity of one, constants have a complexity of two to reduce the excessive use of constants in equations. 
Basic arithmetic operations are considered less complex than non-linear transformations. 
In addition, we employ measures to avoid combinations of operations that would be in agreement with pure dimensional analysis, but not meaningful from a practical perspective.
To this end, we prohibit the nesting of certain functions, such as $\sin(\sin(\circ))$ and other functions.
Tab.~\ref{tab:gp_parameters} outlines these functions and summarizes the previously described algorithm settings. 
Our GP algorithm is implemented in the \texttt{TiSR} framework, which was initially proposed for SR for thermodynamics applications.
However, it is also well-suited to our problem due to its high degree of customizability and integration of state-of-the-art algorithmic components~\cite{martinek_introducing_2023}.
In general, various other SR frameworks could be employed for the SR step of our ML pipeline, such as \texttt{PySR}~\cite{cranmer_interpretable_2023} or \texttt{Operon}~\cite{burlacu_operon_2020}. 
These are, however, more challenging to extend with customized algorithmic components such as the dimensional analysis considering unknown constants, which are essential for our approach.

\begin{table}[t]
    \centering
    \caption{Parameter settings used in the GP algorithm}
    \begin{tabular}{l|l}
    \toprule
    GP parameter & Settings \\
    \midrule
    Population size & 500 \\
    Training duration & 200 generations \\
    Max. complexity of expression & 30 \\
    Optimization objectives & MSE, Spearman correlation, dimension penalty, complexity \\
    HoF objectives & MSE, dimension penalty, complexity \\
    Training features  & $\rbar_j$ $[\unit{\metre}]$, $\theta_j$ $[-]$, $\varphi_j$ $[-]$, $\mathrm{Re_p}$ $[-]$ or $\varepsilon_\mathrm{p}$ $[-]$, tuneable constants $[?]$\\
    Target feature & $\frac{\Delta F_{x,i}}{\langle F_x \rangle}$ $[-]$ \\
    Function set $\mathcal{F}$ & $\{+$, $-$, $\cdot$, $\div$, $\circ^{(\circ)}$, $\sin(\circ)$, $\cos(\circ)$, $e^{(\circ)}$, $\log(\circ)\}$\\
    Complexity values & \\
    \hspace{3mm} Features & 1 \\
    \hspace{3mm} Constants & 2 \\
    \hspace{3mm} $+$, $-$, $\cdot$, $\div$ & 1 \\
    \hspace{3mm} $\circ^{(\circ)}$, $\sin(\circ)$, $\cos(\circ)$, $e^{\circ}$, $\log(\circ)$ & 2 \\
    Prohibited function nesting &  $\sin \rightarrow \{\sin, \cos\}$, $\cos \rightarrow \{\sin, \cos\}$, $\exp \rightarrow \{\exp\}$, \\ 
     & $\log \rightarrow \{\log\}$, $\circ^\circ \rightarrow \{\circ^\circ\}$\\
    Implementation & \texttt{TiSR} \cite{martinek_introducing_2023}\\
    \bottomrule
    \end{tabular}
    \label{tab:gp_parameters}
\end{table}

\subsection{Data preprocessing}
\label{sec:data_preprocess}
We have discussed some data-related preprocessing steps in the previous sections. 
In the following, we summarize how the original data is processed before both training the GNN and the GP algorithm.
We set the number of neighbouring particles considered predicting the target variable to 30.
Related studies suggested diminishing returns in terms of accuracy when more neighbours are considered, so that 30 is a good compromise of accuracy and computational cost~\cite{Siddani2023}. 
In a first step, all relative positions of neighbouring particles are transformed to a spherical coordinate system $(\rbar_j,\theta_j,\varphi_j)$, with the particle of interest located at the origin. $\theta_j$ represents the polar angle relative to the flow direction, while $\varphi_j$ is the azimuthal angle in a plane normal to the flow direction.
The distance $\rbar_j$ is normalized by the particle diameter to facilitate comparison and analysis across different particle sizes and configurations.
Compared to Cartesian coordinates,  this representation simplifies the problem, particularly for the GP algorithm, as it allows the direct application of trigonometric functions to the input features without requiring intermediate transformations. 

Overall, 19,632 particle arrangements are available, which are evenly distributed over the combinations of eight $\mathrm{Re_p}$ and six $\VF$.
Initially, the dataset is split into $70\%$ training and $30\%$ test data. 
Thereby, it is ensured that the distribution of $\mathrm{Re_p}$ and $\VF$ is even among both sets to avoid over- or underrepresentation of certain configurations in the training or testing phase. 
Both sets are augmented by seven incremental rotations of each particle arrangement of $\frac{\pi}{4}$ around the axis of the free stream to cover each quadrant in the coordinate system. 
In this way, the symmetry inherent to the underlying problem is reflected in the datasets, and more data samples are available~\cite{Seyed-Ahmadi2022}.
With the augmentation after the dataset split, it is ensured that the datasets are mutually exclusive, and the test set does not include samples that are rotations of samples from the training set.
Preliminary experiments suggest a slight improvement in accuracy when applying a logarithmic transformation to the Reynolds number, $\log_{10}(\mathrm{Re_p})$.
This transformation can help rescale the feature to make it more suitable for numerical optimization by compressing large variations in $\mathrm{Re_p}$.

During the training of the GNN, $80\%$ of the training data is used for tuning the trainable parameters of the network, and $20\%$ is used as a validation set to detect potential overfitting and trigger early stopping. 
Again, we make sure that these datasets, including their rotations, are mutually exclusive.
The intermediate data products are generated from the samples from the validation datasets.
Since 30 pairwise interactions per configuration are stored, this generates a dataset with 658,560 samples covering pairwise interactions for all $\mathrm{Re_p}$ and $\VF$.
To approximate these pairwise interactions with symbolic expressions, the datasets are filtered by $\VF$ and by $\mathrm{Re_p}$ before the training of the GP algorithm. 
During the training of the GP algorithm, the training data is again split into $80\%$ for evaluating and fitting the constants of the expression, and $20\%$ to trigger early stopping during parameter fitting. 
The test set is a holdout dataset only employed to assess the final performance of the trained models.

\section{Experiments and results}
\label{sec:results}

Using the algorithm settings from Sec.~\ref{sec:algo_settings} and data preprocessing steps as described in Sec.~\ref{sec:data_preprocess}, we conducted experiments for the combination of eight $\mathrm{Re_p}$ and six $\VF$ as introduced in Sec.~\ref{sec:data_gen}.
In the following, we will present and analyse the performance of the GNN and the derived symbolic models. 

\subsection{GNN results}

The GNN training was repeated eleven times, to account for variations in the training process due to the stochastic nature of the parameter initialization.
In this way, we aimed to ensure the reliability of the method and verify that the results do not arise from random chance.
The GNN plays a crucial role in our approach by decomposing the force deviation into pairwise contributions, which reduces the problem complexity tremendously for the subsequent GP step. 
Furthermore, we employ a single GNN model on data spanning all $\mathrm{Re_p}$ and $\VF$, which we expect to improve generalization and to approximate a unified representation of the force deviations.
In the GP step, we then develop separate symbolic models: one for each $\mathrm{Re_p}$ with $\VF$ as a variable and one for each $\VF$ with $\mathrm{Re_p}$ as a variable.
This separation is intended to reduce the search space for the GP algorithm, as previous experiments have shown difficulties in learning accurate equations that remain valid across multiple $\VF$ and $\mathrm{Re_p}$.
We expect that a unified GNN representation, from which the GP training data is extracted, helps to learn symbolic models that are more consistent and structurally similar, despite being trained on separate datasets.
We analyze the performance of the GNN and the symbolic models based on the coefficient of determination, $R^2$. We employ a computation as presented in~\cite{Akiki2017a,Akiki2017}, where $R^2$ is defined as 
\begin{equation}
    R^2 = 1 - \frac{\sum\limits_{i=1}^{N} \left[\Delta F_{x,i}^{\mathrm{true}}  - \Delta F_{x,i}^{\mathrm{model}} \right]^2 }{\sum\limits_{i=1}^{N} \left[ \Delta F_{x,i}^{\mathrm{true}}  - \langle \Delta F_{x}^{\mathrm{true}} \rangle \right]^2 },
    \label{eqn:R2}
\end{equation}
where $\Delta F_{x,i}^{\mathrm{model}}$ and $F_{x,i}^{\mathrm{true}}$ are the predicted and ground truth force deviations of the $i-$th particle in the dataset, respectively. The term $\langle \Delta F_{x}^{\mathrm{true}} \rangle$ represents the mean force deviation, which is expected to be approximately zero since we assume that there exists a perfect mean drag model. The value $R^2$ generally indicates the portion of the force deviations that are captured by the model as compared to the standard deviation of the drag force. A perfect model should achieve $R^2=1$.

\begin{figure}[t]
    \centering
    \includegraphics[height=0.43\linewidth]{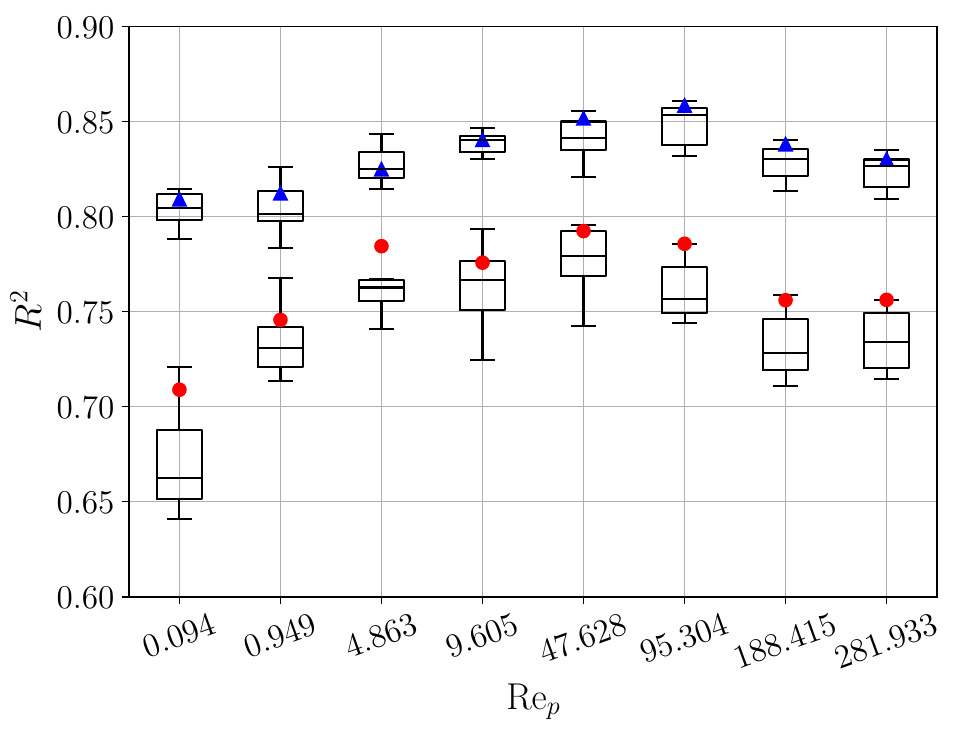}
    \includegraphics[height=0.43\linewidth]{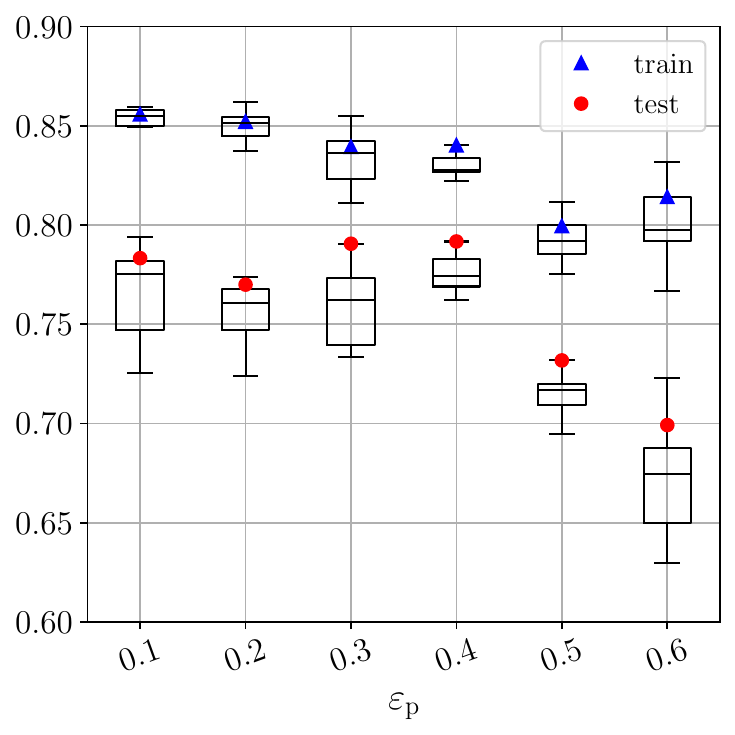}
    \caption{Distribution of $R^2$ over 11 repeated runs of the GNN, disaggregated by $\mathrm{Re_p}$ (left) and $\VF$ (right). The markers indicate the $R^2$ values of the selected model that was used to generate the training data for the GP algorithm.}
    \label{fig:GNNtraintest}
\end{figure}

Fig.~\ref{fig:GNNtraintest} shows the distribution of $R^2$ values across eleven repeated runs of the GNN, disaggregated by the different values of $\mathrm{Re_p}$.
The box plots represent the distributions of $R^2$ scores, showing the median, the interquartile range and the whiskers of each distribution. 
The markers indicate the specific $R^2$ values of the model that was selected as the ``best'' model and used to generate the training data for the GP algorithm:
blue triangles represent the $R^2$ values for the training dataset, red circles represent the $R^2$ values for the test dataset.
Overall, the $R^2$ values are between $0.783$ and $0.861$ on the training set and $0.641$ and $0.799$ on the test set for the different values of $\mathrm{Re_p}$. 
Over the different $\VF$ values, the highest $R^2$ values are $0.880$ on the training and $0.794$ on the test set, and the lowest values are $0.767$ and $0.630$ respectively.
These values are comparable or better than those reported in related studies~\cite{Seyed-Ahmadi2022,Siddani2023}, suggesting that the GNN approach generally achieves a reasonable predictive performance.
By displaying the disaggregated data for $\mathrm{Re_p}$ and $\VF$, we aim to analyze whether a specific parameter configuration is over- or underfitted by the network, despite the equal share of each parameter configuration in the training data.
We observe that the highest $R^2$ scores of up to $0.799$ on the test set are achieved for mid-ranged $\mathrm{Re_p}$ values, i.e., $4.863$, $9.605$, $47.628$, and $95.304$.
At the lower and higher ends of the $\mathrm{Re_p}$ range, the GNN yields lower accuracies and higher distances between train and test sets, with the worst performing GNN having an $R^2$ of $0.641$ for $\mathrm{Re_p}=0.094$.
On the datasets disaggregated by $\VF$, we observe similarly high $R^2$ scores for $\VF$ between $0.1$ and $0.4$. 
For the two highest $\VF$, i.e., $0.5$ and $0.6$, the accuracy on the test set drops, indicating that the GNN struggles to generalize well in denser regimes.

\begin{table}[h]
    \centering
    \caption{$R^2$ scores of the selected GNN model. The entire dataset was disaggregated, once by $\varepsilon_\mathrm{p}$, and once by $\mathrm{Re_p}$, to evaluate the performance for different flow regimes and configurations.}
    \label{tab:R2values}

    \begin{subtable}[c]{0.48\linewidth}
        \centering
        \caption{$R^2$ per $\varepsilon_\mathrm{p}$}
        \begin{tabular}{c|c|c}
            \toprule
            $\varepsilon_\mathrm{p}$ & $R^2$ Train & $R^2$ Test\\ 
            \midrule
            $0.102$ & $0.8558$ & $0.7833$ \\
            $0.200$ & $0.8520$ & $0.7699$ \\
            $0.300$ & $0.8394$ & $0.7906$ \\
            $0.404$ & $0.8402$ & $0.7917$ \\
            $0.500$ & $0.7994$ & $0.7318$ \\
            $0.612$ & $0.8141$ & $0.6992$ \\ 
            \bottomrule
        \end{tabular}
        \label{tab:R2perVF}
    \end{subtable}
    \hfill
    \begin{subtable}[c]{0.48\linewidth}
        \centering
        \caption{$R^2$ per $\mathrm{Re_p}$}
        \begin{tabular}{c|c|c}
            \toprule
            $\mathrm{Re_p}$ & $R^2$ Train & $R^2$ Test\\ 
            \midrule
            $0.094$ & $0.8091$ & $0.7089$ \\
            $0.949$ & $0.8123$ & $0.7457$ \\
            $4.863$ & $0.8249$ & $0.7844$ \\
            $9.605$ & $0.8404$ & $0.7758$ \\
            $47.628$ & $0.8516$ & $0.7924$ \\
            $95.304$ & $0.8584$ & $0.7857$ \\ 
            $188.415$ & $0.8380$ & $0.7560$ \\ 
            $281.933$ & $0.8301$ & $0.7562$ \\ 
            \bottomrule
        \end{tabular}
        \label{tab:R2perRe}
    \end{subtable}

\end{table}


In both plots in Fig.~\ref{fig:GNNtraintest}, we observe a difference between train and test $R^2$ spanning over all $\mathrm{Re_p}$ and $\VF$ values.
However, during the training of the GNNs, the MSE on the validation dataset decreased along with the training MSE, indicating that the model was learning the underlying patterns in the data without overfitting excessively to the training set.
The markers in Fig.~\ref{fig:GNNtraintest} indicate the $R^2$ values of the GNN that were selected among the eleven models for the extraction of the pairwise interactions, which serve a training data for the GP algorithm.
For each of the eleven models, we computed the maximum difference between train and test $R^2$ over the range of $\mathrm{Re_p}$.
The model with the lowest maximum difference was then selected. 
With this, we aimed to select the best generalized model among the trained GNNs, which is an important aspect to develop symbolic models with similar structure. 
The exact numerical $R^2$ values for the selected GNN model can be found in Tab.~\ref{tab:R2values}.
To further investigate the effects of varying $\mathrm{Re_p}$ and $\VF$ on the GNN performance, we analyze the correlation plots obtained by this selected model. 

\begin{figure}
    \centering
    \includegraphics[width=1.0\linewidth]{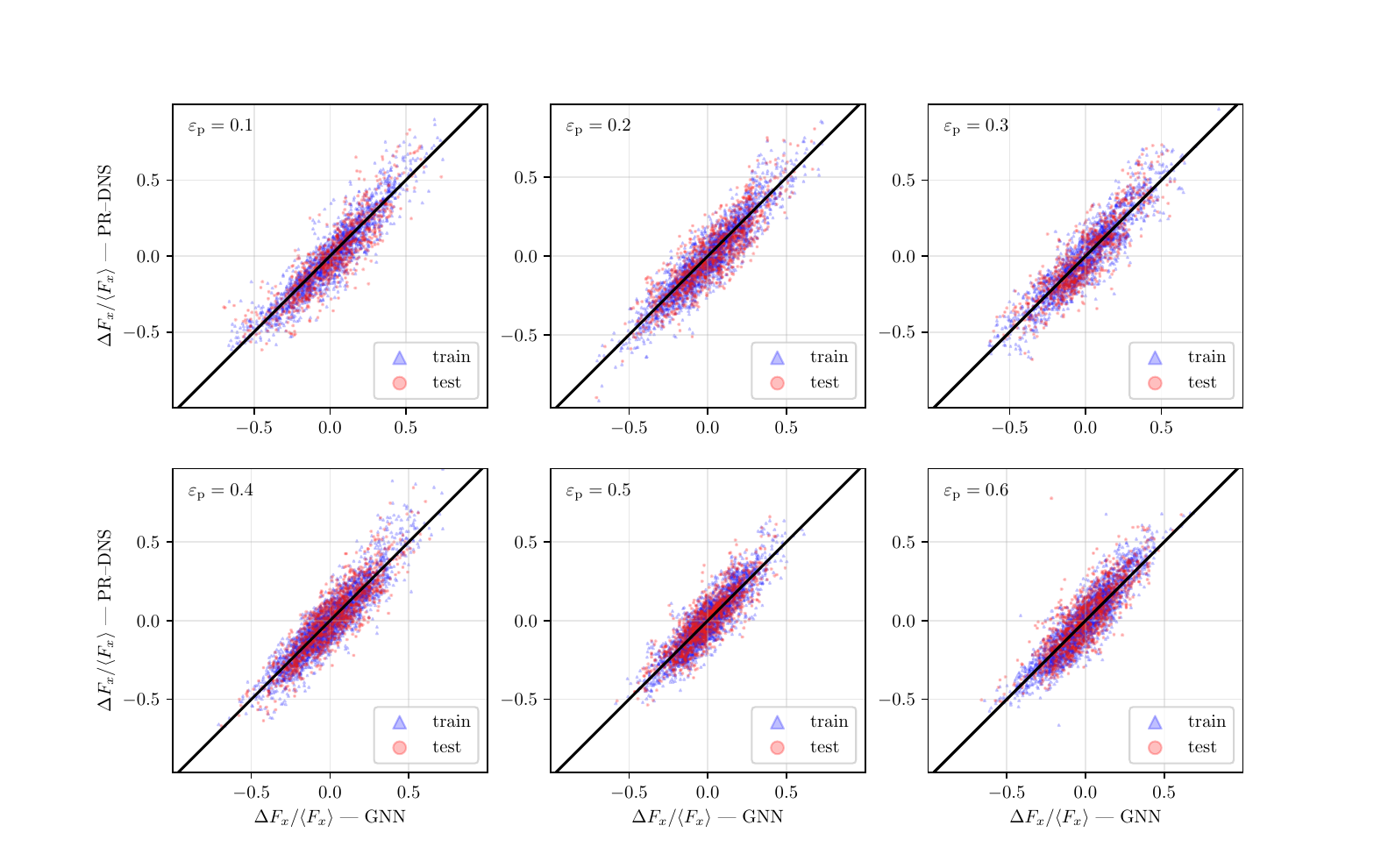}
    \caption{Correlation plots of the predicted deviation from the mean drag as predicted by the GNN (x-axis) and ground truth from the particle-resolved simulation (y-axis) for different $\VF$. Training and test data are shown in blue and red markers, respectively. The diagonal line represents the perfect correlation line.}
    \label{fig:GNN_R2VF}
\end{figure}

\begin{figure}
    \centering
    \includegraphics[width=1.0\linewidth]{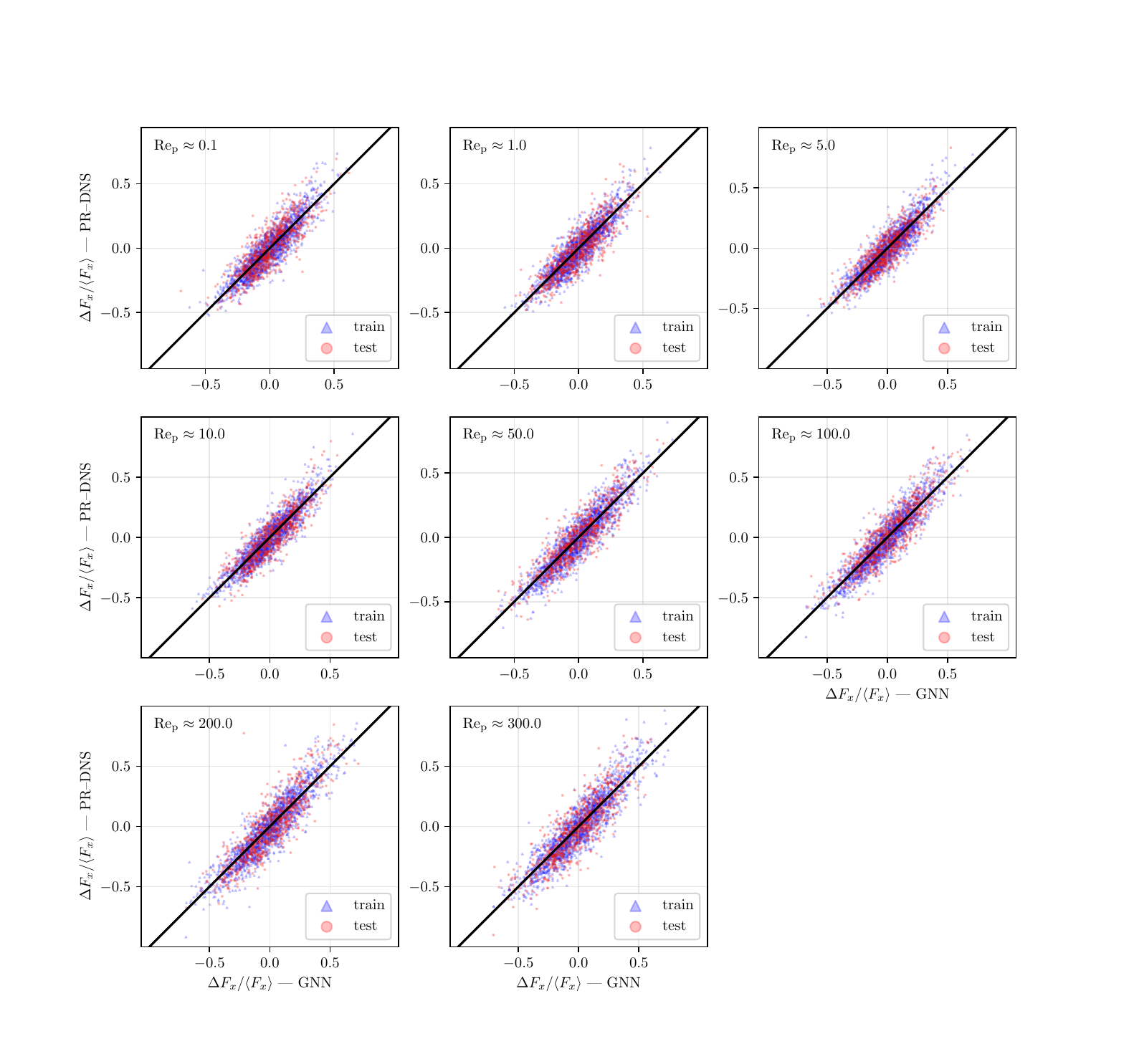}
    \caption{Correlation plots of the predicted deviation from the mean drag as predicted by the GNN (x-axis) and ground truth from the particle-resolved simulation (y-axis) for different $\mathrm{Re_p}$. Training and test data are shown in blue and red markers, respectively. The diagonal line represents the perfect correlation line.}
    \label{fig:GNN_R2RE}
\end{figure}

Fig.~\ref{fig:GNN_R2VF} displays the correlation plots of the model prediction and the particle-resolved simulation
data for different $\VF$, and Fig.~\ref{fig:GNN_R2RE} for different $\mathrm{Re_p}$.
Each marker in the plots represents the prediction result obtained for a single sample of the dataset.
The black line indicates perfect agreement between PR-DNS data and GNN predictions. 
The model demonstrates good correlations with the particle-resolved simulation datasets for both training and test data. 
For those $\VF$ with lower $R^2$ scores, i.e., $\VF=\{0.5, 0.6\}$, we observe more markers with a higher distance to the black line.
Similarly, $\mathrm{Re_p}\approx\{10,50,100\}$ with the highest $R^2$ scores in Tab.~\ref{tab:R2perRe} exhibit markers more concentrated around the black line in Fig.~\ref{fig:GNN_R2RE}. 


\begin{figure}
    \centering
    \includegraphics[width=0.55\linewidth]{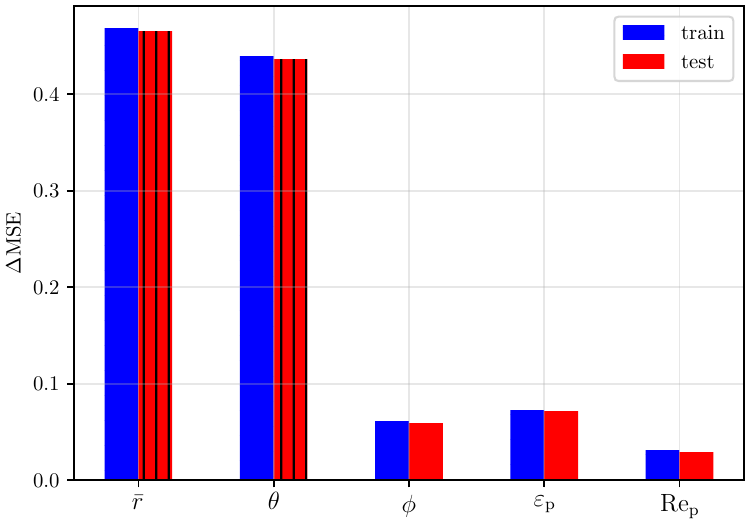}
    \caption{Feature importance analysis of the selected GNN model: the y-axis indicates how much the MSE increases, $\Delta \mathrm{MSE}$, when the feature on the x-axis is permuted. We report $\Delta \mathrm{MSE}$ over ten permutations per feature to eliminate randomness in the shuffling. }
    \label{fig:GNN_MSE_drop}
\end{figure}

As mentioned previously, one significant drawback of purely data-driven ML methods is their lack of interpretability, which limits our understanding of the underlying computations. 
To address this issue, various approaches have been proposed, including the analysis of feature importance. 
This analysis measures the impact on the model predictions or error when one or more features are removed or altered.
We employ a permutation-based feature importance analysis to understand how influential each of the features are in the predictions of the GNN; with this, we moreover gain insights into the features that will most likely be important in the symbolic models, as these are derived from the GNN.
In the permutation-based approach, we shuffle one feature column in the dataset at a time, while keeping the other columns unchanged. 
This column-wise shuffling preserves the overall distribution of feature values compared to the original dataset.
However, it is possible that the permuted dataset may contain combinations of extreme values of multiple features, which were not present in the original data. 
Despite this limitation, this relatively simple method helps to qualitatively estimate how different features contribute to the model predictions. 
To avoid misleading results due to randomness in the shuffling process, we repeat the process ten times for each feature and present the mean increase in terms of MSE in Fig.~\ref{fig:GNN_MSE_drop}.
The datasets for this analysis contain all combinations of $\VF$ and $\mathrm{Re_p}$.

We observe a high importance of $\rbar$ and $\theta$, while the other three features are by orders of magnitude less important. 
In other words, the specific flow regime as a combination of $\mathrm{Re_p}$ and $\VF$ has a lower influence on the model error than the distance and polar angle of the neighbouring particles. 
Since we incorporate the pairwise assumption, we expect the azimuthal angle $\phi$ to not have any influence on the prediction due to rotational symmetry around the flow direction when considering a pair of particles at a time.
This expectation aligns with the equations we previously developed for the Stokes flow, where $\phi$ did not appear in any equation~\cite{Elmestikawy2024}. 
However, the MSE of the GNN slightly increases when the azimuthal angle is permuted, indicating that some influence is present. 
Overall, the importance values for the train and test sets are similar, which further confirms our assumption that no severe overfitting to the training data has occurred. 

Considering the overall good correlation of the GNN predictions with the PD-DNS data (see Figs.~\ref{fig:GNN_R2VF} and~\ref{fig:GNN_R2RE}) as well as the analysis of important features expected to be present in the equations, we extract the pairwise interactions from the selected GNN model and train our proposed GP algorithm to replace the edge model with symbolic expressions. 

\subsection{GP results}
\newcommand{\highlightY}[1]{%
    \colorbox{yellow!30}{$\displaystyle#1$}}
\newcommand{\highlightR}[1]{%
    \colorbox{red!30}{\vrule height15.5pt depth10.5pt width0pt $\displaystyle#1$}}
%
In the following, we report and examine the resulting symbolic models as developed by the GP algorithm. 
As already mentioned, we train the GP algorithm on the pairwise contributions estimated by the GNN. 
For each $\VF$ and $\mathrm{Re_p}$, we repeated the GP algorithm eleven times to account for the non-deterministic nature of the algorithm.
Each run returned a Hall-of-Fame (HoF) with non-dominated solutions each satisfying the HoF objectives, MSE, complexity, and dimension error, to varying extents. 
We first filter the HoFs for equations without unit violations.
Then, we apply the ``top $x\%$'' selection strategy similar to the one proposed in~\cite{de_franca_alleviating_2023} to effectively reduce the number of equations from which we make a final selection. 
For each run, we select the three least complex equations within the top $95\%$ in terms of $R^2$ on the test set. 
The symbolic expressions reported below represent a chosen sample of these expressions that are relatively less complex and achieve reasonable accuracy. 
It should be noted that all the numerical coefficients in the expressions are replaced with an arbitrary coefficient variable, $C_k$, to reveal any existing patterns.\par
\begin{table}
\caption{Sample of GP results when trained on datasets at a given $\mathrm{Re_p}$}
\centering
\begin{tabular}{c|c|r}
\toprule
$\mathrm{Re_p}$ & $R^2$  & Equation \\ \midrule
$0.094$ & $0.5394$ & $\left(C_{2} + e^{\sin{\left(\theta \right)}}\right) \left(C_{0} + \varepsilon_\mathrm{p} \highlightY{\left( C_{1} + \frac{C_{3}}{\bar{r}} \right)} \right)$\\
$0.949$ & $0.6109$ &  $C_{2} \left(C_{0} - \sin^{C_{1}}{\left(\theta \right)}\right) \left(\varepsilon_\mathrm{p} \highlightY{\left( 1 - \frac{C_4}{\bar{r}} \right)} - \frac{C_{3}}{\bar{r}}\right)$\\
$4.863$ & $0.5262$ & $\left(C_{0} + \varepsilon_\mathrm{p} \highlightY{\left( C_1 +  \frac{C_3}{\bar{r}} \right)} + \frac{C_1 \varepsilon_\mathrm{p}^2}{\bar{r}} \right) \left(C_{2} - e^{\sin{\left(C_{4} + \theta \right)}}\right)$\\ 
$9.605$ & $0.5574$ & $\left(C_{5} + \cos{\left(C_{0} \theta + C_{2} \right)}\right) \left(\highlightY{\frac{C_{1}}{\bar{r}} + C_{4}} + C_{3} \varepsilon_\mathrm{p} \left(C_{6} + \bar{r}\right) \right)$\\
$47.628$,& $0.5865$ & $\left(C_{2} - \sin{\left(C_{0} \theta + C_{5} \right)}\right) \left(\frac{C_{1}}{\bar{r}} - \varepsilon_\mathrm{p} \highlightY{\left(C_{3} + \frac{C_{4}}{\bar{r}}\right)}\right)$\\ 
$95.304$ & $0.5903$ &  $\left( \frac{C_0}{\bar{r}} + \varepsilon_\mathrm{p} \highlightY{\left( {C_1} + \frac{C_3}{\bar{r}} \right)} \right) \left(C_{2} + \sin{\left(C_{4} + C_{5} \theta \right)}\right)$\\
$188.415$ & $0.4591$ & $\left(C_{0} + \frac{\sin{\left(C_{3} + C_{4} \theta \right)}}{\bar{r}}\right) \highlightY{\left(C_{1} + \frac{C_{2}}{\bar{r}}\right)}$\\ 
$281.933$ & $0.4943$ & $\bar{r}^{C_{2}} \left(\frac{C_{3} \cos{\left(C_{1} + \theta \right)}}{\bar{r}} + C_{4}\right) \cos{\left(C_{0} + \theta \right)}$ \\ \bottomrule
\end{tabular}
\label{tab:GP_eqn_per_RE}
\end{table}
Table~\ref{tab:GP_eqn_per_RE} shows the results of the GP algorithm when trained with datasets at a given $\mathrm{Re_p}$. By examining the expressions, it is clear that all expressions include $\bar{r}$ and $\theta$, while $\varphi$ never appears in any of them. In the majority of the expressions, $\bar{r}$ has a diminishing impact on the pairwise contribution, i.e., distant neighbours have less pairwise contribution. The exclusion of $\varphi$ is expected as a direct result of incorporating the pairwise assumption in GNN, which is successfully captured by the GP algorithm. The absence of $\varphi$ is also in line with feature importance analysis performed on the trained GNN. Since these expressions result from applying the GP algorithm to datasets of a given $\mathrm{Re_p}$ at a time, expressions cannot have $\mathrm{Re_p}$ as there is no variation in $\mathrm{Re_p}$ across the training samples. Another observation is that $\varepsilon_\mathrm{p}$ appears less frequently than expected. This can be justified by the fact that both $\bar{r}$ and $\varepsilon_\mathrm{p}$ encompass similar characteristics of the arrangement. For denser arrangements of particles, where $\varepsilon_\mathrm{p}$ is high, $\bar{r}$ of the nearest neighbours is relatively lower and vice versa.\par
Repeated patterns can be recognized in the GP expressions, some of which are highlighted in yellow. These patterns and sub-expressions can potentially be used to tailor a symbolic expression that mimics the behaviour of the GNN, while satisfying meaningful constraints \cite{Elmestikawy2024}. Although the GP expressions are compact and traceable, there is still a gap in terms of predictive capabilities between the GP expression and the trained GNN. This is evident by comparing the $R^2$ values reported in Table~\ref{tab:R2perRe} for the GNN and their counterparts of the symbolic models in Table~\ref{tab:GP_eqn_per_RE}.\par
\begin{table}
\caption{Sample of GP results when trained on datasets at a given $\varepsilon_\mathrm{p}$}
\centering
\begin{tabular}{c|c|r}
\toprule
$\varepsilon_\mathrm{p}$ & $R^2$ & Equation \\ \midrule
$0.1$ & $0.6359$ & $ \  \left(C_{1} - \sin{\left(C_{5} + \frac{C_{0} - \theta \left(C_{3} + \bar{r}\right)}{\bar{r}} \right)}\right) \highlightY{\left(C_{2} + \frac{C_{4}}{\bar{r}}\right)}$ \\
$0.2$ & $0.6269$ & $ \  \bar{r}^{C_{3}} \left(C_{0} + \frac{C_{2} \theta}{C_{4} + \mathrm{Re_p}}\right) \left(C_{1} - \cos{\left(\theta \right)}\right)$ \\
$0.3$ & $0.5806$ & $ \  \left(C_{3} - \sin{\left(C_{2} + C_{4} \theta \right)}\right) \highlightY{\left(\frac{C_{0}}{\bar{r}} + C_{1}\right)}$ \\
$0.4$ & $0.5226$ & $ \  \highlightY{\left(C_{1} + \frac{C_{3}}{\bar{r}}\right)} \left(C_{2} + \cos{\left(C_{0} \theta \right)}\right)$ \\
$0.5$ & $0.5237$ & $ \  \highlightY{\left(C_{0} + \frac{C_{3}}{\bar{r}}\right)} \left(C_{4} \bar{r} + \sin{\left(C_{1} \theta + C_{2} \right)}\right)$ \\
$0.6$ & $0.5080$ & $ \  \highlightY{\left(\frac{C_{1}}{\bar{r}} + C_{4}\right)} \left(C_{0} + \frac{C_{3}}{\bar{r}} - \cos{\left(C_{2} \theta \right)}\right)$ \\\bottomrule
\end{tabular}
\label{tab:GP_eqn_per_VF}
\end{table}
In Table~\ref{tab:GP_eqn_per_VF}, a sample of the expressions generated by the GP algorithm when trained on datasets at a given $\varepsilon_\mathrm{p}$ is presented. Similar to the results in Table~\ref{tab:GP_eqn_per_RE}, these expressions exhibit recurring patterns. These patterns could serve as building blocks for constructing more complex and generalized symbolic models. $\bar{r}$ and $\theta$ remain prevalent in all expressions. Contrary to expectations, $\mathrm{Re_p}$ does not appear frequently in the expressions generated by the GP algorithm, although $\mathrm{Re_p}$ is allowed to vary in the considered datasets. This is, however, consistent with the feature importance study conducted on the GNN and confirms that the GP symbolic models, to some extent, replicate the behaviour of the GNN.\par
The $R^2$ values in Table~\ref{tab:GP_eqn_per_VF} indicate the same gap in predictive accuracy of the GP expressions when compared against the GNN performance as reported in Table~\ref{tab:R2perVF}. $R^2$ decreases as $\varepsilon_\mathrm{p}$ increases, which is a trend observed in the GNN performance as well. Decrease in accuracy for denser particle arrangements is expected generally for models that rely on pairwise assumption, as the assumption loses its validity in denser arrangements~\cite{Akiki2017}. Overall, the GP results demonstrate the potential to uncover interpretable symbolic models for the drag variation. However, further refinement and incorporation of physical and/or mathematical constraints are necessary to limit the search space for the GP algorithm. 

%
\section{Summary and conclusions}
\label{sec:conclusion}
In this paper, we extend our previous work on deterministic drag modelling using a machine learning framework to higher Reynolds number regimes and denser particle volume fractions. Using particle-resolved direct numerical simulation (PR-DNS) data, we train a graph neural network (GNN) to estimate the pairwise contribution to the drag variation. Then a genetic programming~(GP) algorithm is employed to derive drag deviation models in the form of symbolic expressions. The GNN achieves a coefficient of determination, $R^2$, comparable to those reported in related studies, despite training a single network for all flow regimes and volume fractions. This demonstrates the robustness and predictive capabilities of the GNN, as it successfully captured the drag variation across a wide range of Reynolds numbers $\mathrm{Re_p}$ and volume fractions $\varepsilon_\mathrm{p}$.\par
A feature importance analysis is conducted on the trained GNN using the method of feature permutation. This analysis indicates that the normalized inter-particle distance, $\bar{r}$, and the polar angle $\theta$ of the neighbors have more influence on the drag deviations compared to the azimuthal angle, $\varphi$. The analysis also indicates that the global flow conditions $(\mathrm{Re_p},\varepsilon_\mathrm{p})$, usually used in mean drag models, are relatively less impactful on the drag deviations of each individual particle.  This suggests that the local arrangement of particles plays a critical role in drag variation, while the global flow conditions have a secondary effect. \par
The GP algorithm generated symbolic models with lower $R^2$ values compared to the GNN, although GP was applied on datasets of a constant $\mathrm{Re_p}$ and $\varepsilon_\mathrm{p}$ separately. When applied to the constant $\varepsilon_\mathrm{p}$ datasets, symbolic models showed almost no dependency on $\mathrm{Re_p}$, which is in-line with the low feature importance of $\mathrm{Re_p}$ in GNN. This discrepancy highlights a limitation of the current approach and suggests the need for future research to better account for global input features. One potential solution is the multi-view approach~\cite{russeil_multiview_2024}, where separate constants in the equations for each Re and VF are fitted during the training of the GP algorithm.\par
This study represents the first attempt to develop symbolic models for determining the drag variations on particles in assemblies with higher $\mathrm{Re_p}$ and $\varepsilon_\mathrm{p}$. Despite the varying flow conditions, recurring patterns and sub-expressions can be identified in the GP results. These patters and sub-expressions can serve as the foundation for developing a generalized symbolic drag deviation model that applies across different $(\mathrm{Re_p},\varepsilon_\mathrm{p})$.\par
Further refinements to the symbolic regression approach can be achieved by restricting the input features from being arguments to certain GP building blocks, thereby limiting the search space. Additionally, incorporating shape constraints, as introduced in Sec.~\ref{sec:gp}, could ensure that the symbolic models adhere to physical limits, such as the vanishing effect of neighboring particles as the distance $\bar{r}$ approaches infinity. These constraints would not only enhance the robustness of the models, but also reduce the complexity of the search space for the GP algorithm.\par
In conclusion, this work demonstrates the potential of combining GNNs and symbolic regression to develop interpretable drag models for particle-laden flows. While the GNN provides high accuracy, the symbolic models offer a potential insight into the underlying physics and can be more easily integrated into simulation frameworks. Future work should focus on further refining the search space for the GP algorithm. This study paves the way for further advancements in the development of interpretable and accurate drag models for complex fluid-particle systems.
\section{Acknowledgments}
\label{sec:acknowledgments}
The authors acknowledge the funding by the Deutsche Forschungsgemeinschaft (DFG, German Research Foundation) under project number $466092867$ within the Priority Programme ``SPP 2331: Machine Learning in Chemical Engineering''. 
They also acknowledge the funding by the German Federal Ministry of Education and Research
through the 6G-ANNA project (grant no. 16KISK092).
They furthermore acknowledge the computing time granted on the HPC-cluster Sofja at OvGU Magdeburg. 

\appendix


\end{document}